\documentclass[prd,amsmath,amssymb,floatfix,superscriptaddress]{revtex4}
\usepackage{graphicx}
\usepackage{hyperref}
\usepackage{bm}

\begin{document}

\title{Long-Lived Time-Dependent Remnants During Cosmological Symmetry
Breaking: From Inflation to the Electroweak Scale}

\author{Marcelo Gleiser}
\email{mgleiser@dartmouth.edu}
\affiliation{Department of Physics and Astronomy, Dartmouth College,
Hanover, NH 03755, USA}

\author{Noah Graham}
\email{ngraham@middlebury.edu}
\affiliation{Department of Physics, Middlebury College,
Middlebury, VT 05753, USA}

\author{Nikitas Stamatopoulos}
\email{nstamato@dartmouth.edu}
\affiliation{Department of Physics and Astronomy, Dartmouth College,
Hanover, NH 03755, USA}

\date{\today}

\begin{abstract}
Through a detailed numerical investigation in three spatial
dimensions, we demonstrate that long-lived time-dependent field
configurations emerge dynamically during symmetry breaking in an
expanding de Sitter spacetime. We investigate two situations: a single
scalar field with a double-well potential and an
SU(2) non-Abelian Higgs model. For the single scalar, we show that
large-amplitude oscillon configurations emerge spontaneously and
persist to contribute about 1.2\% of the energy density of the
universe. We also show that for a range of parameters, oscillon
lifetimes are {\it enhanced} by the expansion and that this effect is
a result of  parametric resonance. For the SU(2) case, we see about
$4\%$ of the final energy density in oscillons.
\end{abstract}

\maketitle

\section{Introduction}
Spontaneous symmetry breaking plays a key role in our current
understanding of particle physics and is expected to have been a major
factor in determining the physical properties of the early
universe \cite{Vilenkin}. In cosmology, two aspects of symmetry
breaking are of great
interest: it typically happens far from thermal equilibrium and
it is inherently nonlinear. In the context of the
electroweak phase transition, for example, an initially thermalized
state is tossed out of equilibrium as the Higgs evolves to acquire a
nonzero expectation value. In inflation, a nonthermal state
thermalizes to reheat the universe with an explosive energy transfer
from the inflaton to other field modes. It is thus of great interest
to study the dynamics of symmetry breaking in an expanding
background numerically in order to isolate key features that may
escape analytical techniques.

Here, we report results on 3d simulations for two
situations: a single, self-interacting scalar field with a double-well
potential, and an SU(2) non-Abelian Higgs model. In
Refs. \cite{graham_cos,farhi_cos} results have been obtained for the
case of a single scalar in 1d. It was shown that long-lived,
time-dependent field configurations known as oscillons
\cite{bogol,gleiser,copeland} emerged spontaneously and contributed an
amazing 50\% of the total energy density. These initial results
triggered the present study in the context of more realistic
models. There are two broad classes of scalar field
oscillons that have been studied in the literature, small and
large-amplitude. Small-amplitude oscillons do not probe the 
highly nonlinear domain of the potential, and typically have large
spatial widths \cite{fodor,farhi_cos,amin}. Their small amplitude
makes it possible to study them using linearization
techniques. Large-amplitude oscillons are harder to investigate
analytically \cite{dosc,sicilia,hertzberg}. Simulations of scalar
models in static 2d and 3d backgrounds \cite{oscnuc} and expanding 1d
backgrounds \cite{graham_cos,farhi_cos}
indicate that mostly large-amplitude oscillons are excited during
symmetry breaking. As we show next, this is also the case for an
expanding 3d spacetime. The situation is different for SU(2)
models, as we explain below.

This paper is organized as follows: in the next section, we introduce
the scalar field model in an expanding universe and discuss its
lattice implementation. We report our results for a double well
potential, showing that oscillons contribute about 1.2\% of the energy
density. In section III we show that, contrary to naive expectation,
for certain values of the expansion rate oscillons may have their
lifetimes enhanced. We explain this result analytically by making use
of parametric resonance. In section IV we introduce the SU(2)
non-Abelian Higgs model and discuss its lattice implementation in an
expanding universe. In section V we discuss the results for this
model. In particular, we show that, as in the case of a real scalar
field, oscillons contribute a nontrivial percentage of the total energy
density. Furthermore, our results indicate that the cosmological
expansion seems to favor the formation of oscillons for a wider range
of parameters as compared to the static case, where oscillons were
found only in a 2:1 mass ratio for the Higgs and gauge boson. In
section VI, we briefly discuss possible application of oscillons in
cosmology, which we hope to explore in forthcoming work, and
conclude with a summary of our results.

\section{Scalar Field Model}
We consider a scalar field $\Phi({\bf x},t)$ propagating in
$(3+1)$-dimensional de Sitter spacetime with Hubble constant $H = \dot
a/a$ and a double-well potential $
V(\Phi)=(\lambda/4)[\Phi^2-\mu^2/\lambda]^2$. Using $\hbar=c=k_B=1$
and defining dimensionless variables $\phi =
\Phi(\mu/\sqrt{\lambda})^{-1}$ and $\tilde{x}^{\nu} = \mu
x^{\nu}~(\nu=0,1,2,3)$, the equation of motion satisfied by $\phi$ is
\begin{equation}
\ddot{\phi}+3 \frac{\dot{a}(t)}{a(t)}\dot{\phi}=
\frac{\nabla^2\phi}{a(t)^2}+\phi-\phi^3,
\label{Eq:EOM3D}
\end{equation}
where overdot and $\nabla$ denote derivatives with respect to
dimensionless time $\tilde{x}^0$ and space $\tilde{x}^i$.
The expansion rate becomes $H = \mu \tilde{H}$, where
$\tilde{H}\equiv d\ln(a)/d\tilde{x}^0$.

Our initial conditions simulate quasi-thermal states of the free
massive scalar field. The parameters that control the distribution of
the lattice modes are the temperature $T$ and the mass of the field
$m=\mu\sqrt{2}$. The simulation space consists of a cube with comoving
size $L$ and volume $V=L^3$ discretized on a regular lattice with
spacing $\Delta x^i=\Delta r~(i=1,2,3)$. We apply periodic
boundary conditions and label the free field's normal modes by
$\mathbf{k}=(2\pi \mathbf{n}_i/L)$, where $\mathbf{n}=(n_x,n_y,n_z)$
and the $n_i$ are integers $n_i=-N/2+1\ldots N/2$. 
Here $N=L/\Delta r$ is the
number of lattice points per side. Each free mode is
described by a harmonic oscillator with frequency
$\omega_k^2=(2\sin(k\Delta r/2)/\Delta r)^2+m^2$,
where $k=|\mathbf{k}|$. The initial conditions for the field $\phi$
are then given by 
\begin{eqnarray}
 \phi(\mathbf{r},t=0)&=&\frac{1}{\sqrt{V}}
\sum_\mathbf{k}\sqrt{\frac{\hbar}{2\omega_k}}\left[\alpha_ke^{i
\mathbf{k}\cdot\mathbf{r}}+
\alpha_k^*e^{-i \mathbf{k}\cdot\mathbf{r}}\right],\nonumber\\
\dot{\phi}(\mathbf{r},t=0)&=&\frac{1}{\sqrt{V}}
\sum_\mathbf{k}\sqrt{\frac{\hbar\omega_k}{2}}
\left[\alpha_ke^{i \mathbf{k}\cdot\mathbf{r}}-
\alpha_k^*e^{-i \mathbf{k}\cdot\mathbf{r}}\right],
\label{Eq:InitialCond}
\end{eqnarray}
where $\alpha_k$ is a random complex variable with phase distributed
uniformly on $[0,2\pi)$ and magnitude drawn from a Gaussian
distribution such that
$\langle|\alpha_k|^2\rangle=[\coth(\hbar\omega_k/2T)-1]/2$.
This is the amplitude distribution for a quantum harmonic oscillator
\cite{Landau} with the zero-point motion subtracted. On average, modes
with $\hbar\omega_k\lesssim T$ get assigned energy $T$, in agreement
with equipartition, while the energy per mode goes
rapidly to zero for $\hbar\omega_k\gtrsim T$. We thus need a
lattice fine enough to resolve the high $k$ modes that
are excited at high temperatures. Using a value of $\Delta r_0$ that
is at least 10 times smaller than the wavelength of the mode
satisfying $\hbar\omega_k\sim T$ is enough to provide a good continuum limit.

We discretize the equation of motion using second-order space
derivatives with lattice spacing $\Delta r$ in all directions. We then
step forward in time using a fourth-order R\"{u}nge-Kutta method. 
By the Courant condition, we need to keep $\Delta t<a(t)\Delta r$ at
all times. We impose a maximum physical lattice spacing $\Delta
r_{\rm max}$ that is fine enough to resolve field configurations at
physical sizes that we expect for oscillons. When $a(t) \Delta r
\ge \Delta r_{\rm max}$, we refine the lattice by bringing the lattice
spacing back to $\Delta r_{\rm max}/2$ and inserting points by polynomial
interpolation. We pick $\Delta r_{\rm max}$ and $\Delta t$ small
enough so that any further reduction does not significantly affect the
final configuration of a run. All our simulations maintain energy
conservation to a part in $10^3$ or better.

\begin{figure}[htbp]
\includegraphics[scale=0.4]{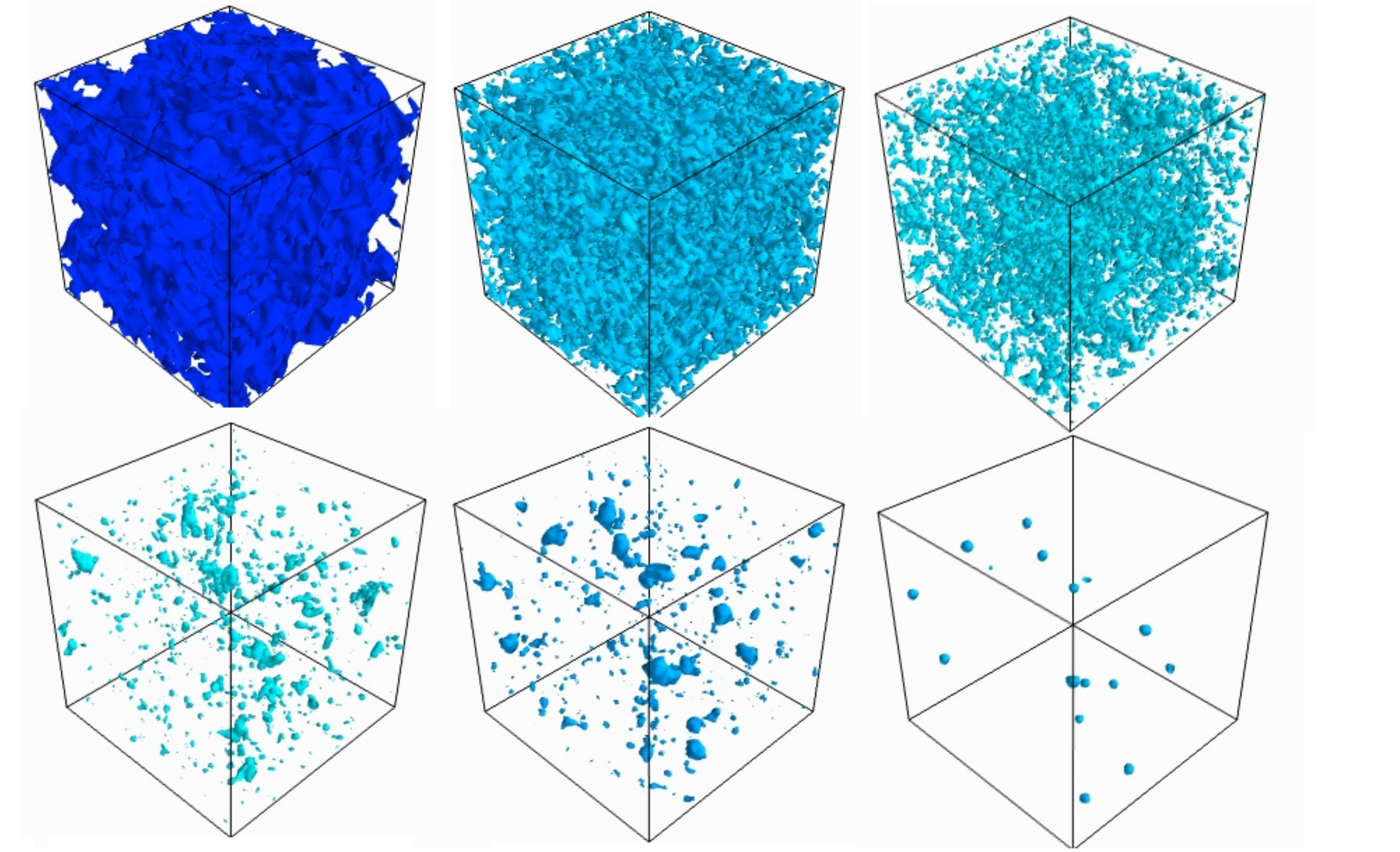}
\caption{Sequence of time snapshots of the energy density. Time
increases from left to right and top to bottom at times
$t\mu={0,50,100,150,200,250}$, $H=0.01\mu$ and $T=6.0\mu$. All
snapshots show the energy density $u$
with an isosurface at $u=0.2\mu^4$.}
\label{Fig:Snapshots}
\end{figure}

We evolve the field $\phi(\mathbf{x},t)$ in a box with $256^3$ lattice
points and $\Delta r_0=0.05 \mu^{-1}$. We keep $\Delta t=0.01\mu^{-1}$
constant throughout the simulation. As the lattice spacing increases
to $\Delta r_{\rm max}=0.5\mu^{-1}$, we insert points in the lattice,
bringing the spacing down to $\Delta r_{\rm max}/2$. We vary the
values of the expansion factor $H$ and the initial temperature $T$
and evolve the field until it cools down to $T/a(t)=0.3\mu$. After the
universe has expanded and cooled, we observed persistent localized
structures as peaks in the energy density. We show a typical sequence
of snapshots in  Fig.~\ref{Fig:Snapshots}.

\begin{figure}[htbp]
\includegraphics[scale=0.4]{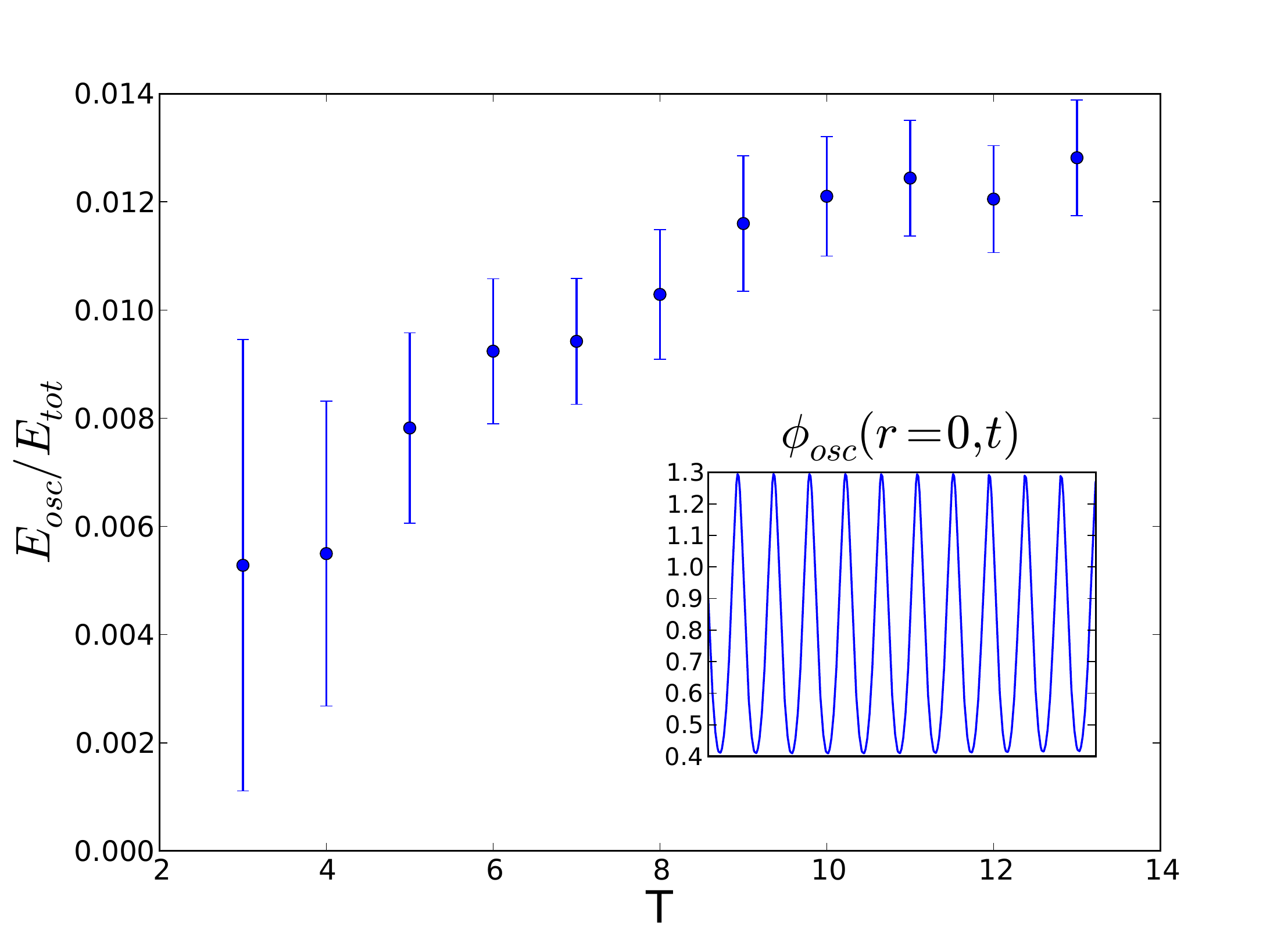}
\caption{Fraction of energy in oscillons as a function of
temperature in units of $\mu$. Here $\Delta r_0=0.05 \mu^{-1}$,
$\Delta r_{\rm max}=0.5 \mu^{-1}$, and $H=0.01 \mu$. Error bars denote
ensemble averages over 10 runs. The inset shows the near-harmonic
oscillations of the oscillon core. A simulation of a typical run can
be viewed at \href{http://www.youtube.com/watch?v=_0co05XkNMY}
{http://www.youtube.com/watch?v=\_0co05XkNMY}
(web link on electronic version). }
\label{Fig:Energy_Fraction}
\end{figure}

By isolating these peaks
individually, we find that they all share the typical signatures of
spherically-symmetric oscillon configurations: their centers
oscillate with the typical oscillon frequency, as shown in the inset of
Fig.~\ref{Fig:Energy_Fraction}, and their energies coincide with
the plateau energies found in detailed oscillon studies
\cite{gleiser,copeland,sicilia}.
The energy of a configuration is calculated by integrating the field's
total energy around its peak using a radius $r=10\mu^{-1}$. We
consistently found $E_{\rm osc} \simeq 45\mu/\lambda$. We then
measured the fraction of energy in oscillons (Fig.\
\ref{Fig:Energy_Fraction}) and the number of oscillons nucleated as a
function of temperature, which scales simply as $N_{\rm osc} \propto
V=L^3 \propto T^3$.

Although we quote results for $H=0.01\mu$, we have performed simulations 
for a slower expansion rate of $H=0.005 \mu$
obtaining similar qualitative behavior:  for a wide range of initial
temperatures, $\rho_{\rm osc}/\rho_{\rm tot} =\Omega_{\rm osc} \sim
1.2\%$. Smaller values of $H$ require impractical computation time,
but we don't expect any qualitative changes. We note that since the
simulations end with fairly large values of $\Delta r=0.5\mu^{-1}$, our
results are lower bounds on $\rho_{\rm osc}$. Of course, scalar field
oscillons are not stable in 3d and will decay after $\tau_{\rm osc} \sim
10^4\mu^{-1}$. Nevertheless, during their lifetime, they may be
responsible for several important effects, as we discuss in section VI. We
also note that oscillons are prevented from forming if the horizon
size $1/H$ is of the order of the oscillon size,
$R_{\rm osc} \sim 4\mu^{-1}$. For
$H\gtrsim0.1\mu$, which fortunately is not very realistic,
large-amplitude fluctuations are flattened out before the stabilizing
effect of nonlinearities can kick in. In other words, for oscillons to
be cosmologically viable, we must have
$R_{\rm osc}/\lambda_H = \tilde{R}_{\rm osc}\tilde{H} \ll 1$,
where $\lambda_H = H^{-1}$ is the horizon length. This condition is
easily satisfied for physics below the Planck scale.

\section{Lifetime Enhancement}
Having established that oscillons emerge dynamically in an expanding
background, we need to examine how the expansion affects their
lifetime. For numerical efficiency, we exploit the spherical symmetry
of the final oscillon configuration and
reduce our system to an effectively 1d problem by letting
$\nabla^2\phi\rightarrow\partial^2\phi/\partial
r^2+(2/r)\partial\phi/\partial r$ in Eq.~\ref{Eq:EOM3D}. 
We find oscillons by setting the initial field configuration to be
Gaussian, $ \phi(r,0)=2\exp(-r^2/R_0^2)-1$, with
boundary conditions $\phi(r\rightarrow\infty,t)=-1\textrm{,
}\phi'(0,t)=0\textrm{, and } \dot{\phi}(r,0)=0$
\cite{gleiser,copeland}. In the absence of expansion, Gaussians with
$2.4\lesssim R_0 \mu\lesssim 4.5$ settle into long-lived oscillon
configurations.

\begin{figure}[htbp]
\includegraphics[scale=0.4]{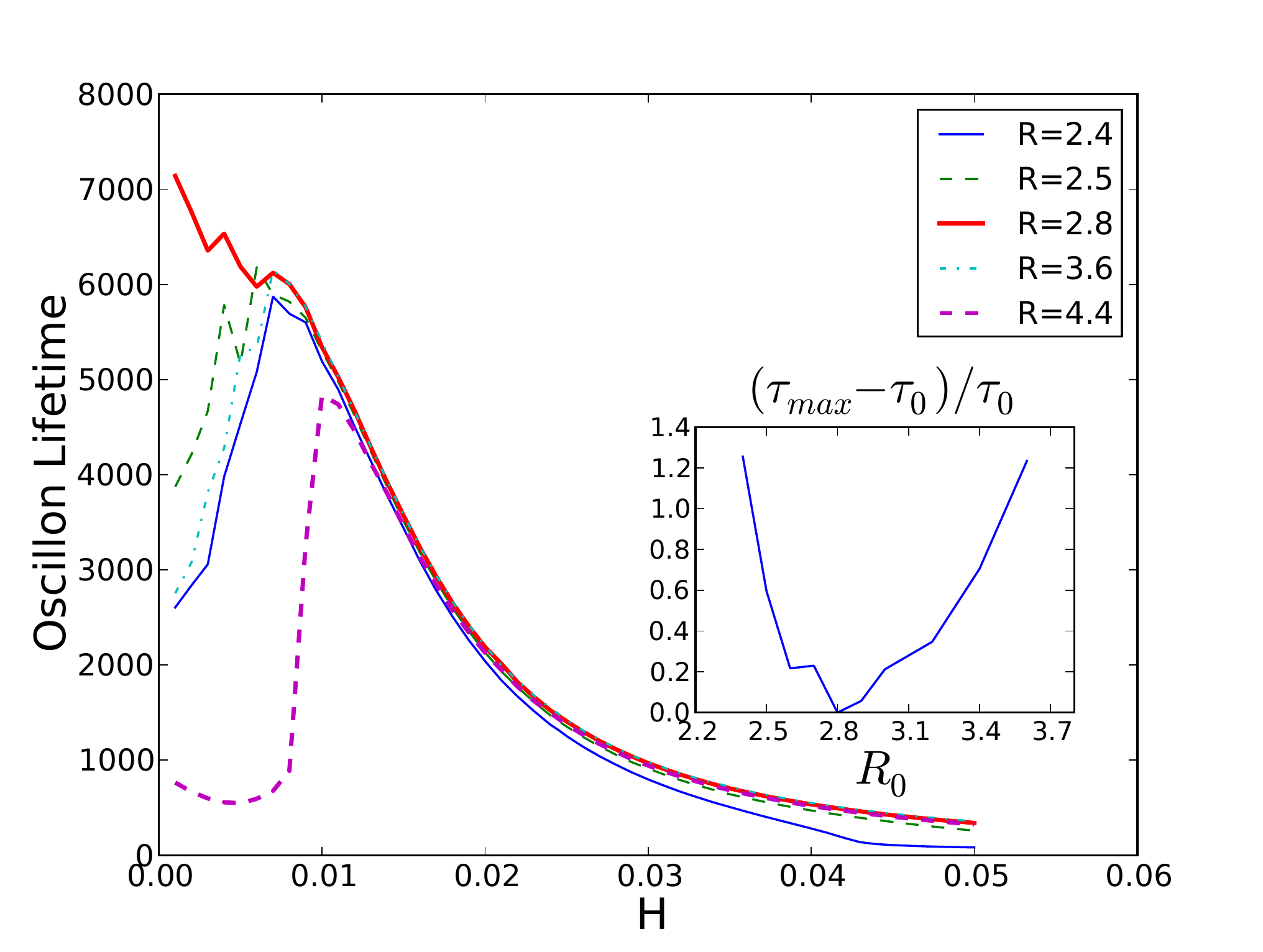}
\caption{Lifetime, in units of $\mu^{-1}$, for oscillons formed from
Gaussians with $2.4\leq R\mu\leq4.4$, as a function of expansion
rate $H$ in units of $\mu$. The inset shows the maximal fractional
increase in oscillon lifetime for different radii at $H_{\rm max}$.
}
\label{Fig:Enhancement}
\end{figure}

We follow the same procedure as in 3d so that as soon as the lattice
spacing becomes $\Delta r_{\rm max}=0.1\mu^{-1}$, 
$L_{\rm max}\gtrsim 2/H$, we insert 
points via polynomial interpolation, and bring the lattice spacing
back to $\Delta r=0.05\mu^{-1}$. We then truncate the box to
$L\gtrsim 1/H$, which can't affect the oscillon at $r=0$. We always use
a box of initial size $L_0=1/H+50\mu^{-1}$ in natural units, and we have
verified that any run with $L_0\gtrsim 1/H$ gives identical results. In
Fig.~\ref{Fig:Enhancement} we show the effects of expansion for a
sample of initial configurations. There is a clear symmetry about
$R_0=2.86\mu^{-1}$, the longest-lived oscillon in the absence of expansion:
radii to both sides of $R_0=2.86\mu^{-1}$ experience an increase in lifetime
for a range of $H$, with the increase being more pronounced for shorter
lifetimes. The longest-lived oscillon, in turn, doesn't experience any
noticeable enhancement. The inset of
Fig.~\ref{Fig:Enhancement} shows the maximum fractional
increase in lifetime $(\tau_{\rm max} - \tau_0)/\tau_0$ as a function
of initial radius $R_0$. The lifetime enhancement follows an approximate
scaling law around $R_0=2.86\mu^{-1}$, $\tau_{\rm osc}\mu 
\sim |R_0\mu - 2.86|^{0.05}$.

To understand the origin of the lifetime enhancement caused by
the expansion, we decompose the field as
$\phi(\textbf{x},t)=\phi_{av}(t)+\delta\phi(\textbf{x},t)$, where
$\phi_{av}$ is the volume averaged field. Linearizing
Eq.~\ref{Eq:EOM3D} with respect to $\delta\phi(\textbf{x},t)$ and
taking the Fourier transform, we obtain (for $k>0$)

\begin{equation}
\delta\ddot{\phi}(k,t) + 3H\delta\dot{\phi}(k,t)+
\left(\frac{k^2}{a(t)^2} + V''(\phi_{av}(t))\right)\delta\phi(k,t)=0.
\label{Eq:LinearEq}
\end{equation}
Once the Gaussian has settled into the oscillon stage,
$V''(\phi_{av}(t))$ can be approximated by
$V''(\phi_{av}(t))=\Phi_0\cos(\omega t)+C$, where $\Phi_0$ and $C$
vary very slowly during an oscillon's lifetime and depend on the value
of the initial radius $R_0$ and the Hubble constant $H$. Here $\omega<m$ is
the oscillon's frequency of oscillation. 
Introducing new variables $\omega t=2z-\pi$, and
$\delta\phi=\exp(-3Hz/\omega)\chi$, Eq.~\ref{Eq:LinearEq} becomes
\begin{equation}
 \chi''+\left[A_k-2q\cos2z\right]\chi=0,
\label{Eq:Mathieu}
\end{equation}
where $A_k=\frac{1}{\omega^2}\left[4k^2/a^2+4C-9H^2\right]$,
$q=2\Phi_0/\omega^2$, and prime denotes differentiation with respect
to the new variable $z$. Eq.~\ref{Eq:Mathieu} is the Mathieu equation,
which is known to exhibit parametric resonance when $A_k\simeq l^2$,
$l=1,2,\ldots$\cite{mathieu_inflation}. Thus, particular combinations
of values of $C$, $H$ and $\omega$ can lead to exponential
amplification in the oscillations of $\chi$, and consequently
$\delta\phi(k,t)$, at certain modes $k/a$. Because $C$ is a positive
number, the $A_k\simeq1$ resonance window occurs for real values of
$k/a$ only for very large values of $H$, which destabilize the oscillon
before it can lead to resonance. The $A_k\simeq 9$ and higher windows
lead to resonances that are too weak to overcome the damping due to
the expansion. The $A_k\simeq 4$ window, however, can lead to
parametric amplification of the dominant oscillon wavevectors for the
values of $H$ that generate the observed lifetime enhancement depicted
in Fig.~\ref{Fig:Enhancement}.

\begin{figure}[htbp]
\includegraphics[scale=0.4]{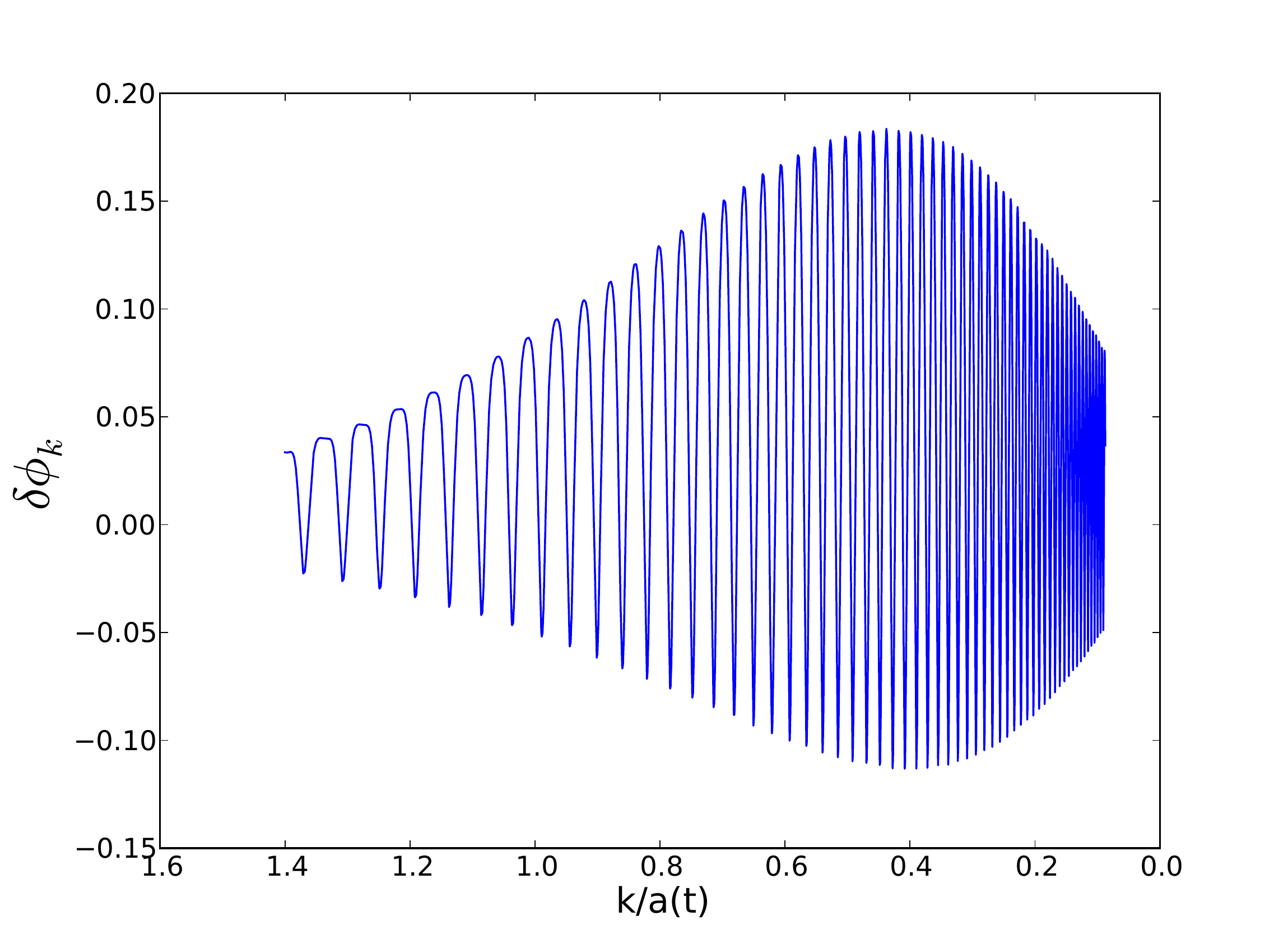}
\caption{Oscillations in $\delta\phi(k,t)$ for an oscillon with
$R_0=2.5\mu^{-1}$ and expansion rate $H=0.006\mu$, at times 
$3200\leq t \mu\leq 3700$. Here
we plot a mode which starts at $k/a\simeq 1.4\mu$ and gets redshifted to
$k/a\simeq 0.1\mu$. As $k/a$ gets smaller, the mode enters the
resonance window, and its Fourier component gets amplified and then
redshifted away. In this case, $\omega\simeq 1.4 \mu$ ($<m=\sqrt{2}
\mu$) and $C=1.8 \mu^2$, with $A_k=4$ for $k/a\simeq0.4 \mu$, which is
the dominant wave vector of this oscillon ($R\simeq 2.9\mu^{-1})$.}
\label{Fig:Resonance}
\end{figure}

The $A_k\simeq4$ resonance window leads to exponential amplification
to the oscillations in $\chi\propto e^{\xi z}$, where 
$\xi \simeq\sqrt{5}q^2/48$ \cite{mathieu_functions}.
For small values of $H$, the amplification overcomes the damping due
to expansion. For higher values of $H$, the damping overcomes the
amplification, and the oscillon decays. An example is shown in
Fig. \ref{Fig:Resonance}.

\section{SU(2) Model: Implementation}
Because the SU(2) model is considerably more expensive numerically to
simulate, the range of experiments we can carry out is
limited. However, these experiments show very similar behavior
to the scalar model. We begin from the Lagrangian density in the
absence of expansion,
\begin{equation}
{\cal L} =  -\frac{1}{4} {\bm F}_{\mu \nu} \cdot {\bm F}^{\mu \nu}
+ (D_\mu \Phi)^\dagger (D^\mu \Phi)
- \lambda(|\Phi|^2 - \mu^2)^2 \,,
\end{equation}
where the boldface vector notation refers to isovectors. Here 
the Higgs field $\Phi$ is an SU(2) doublet, and the SU(2) field
strength and covariant derivatives are
\begin{eqnarray}
\bm{F}_{\mu \nu} &=& \partial_\mu {\bm W}_\nu - \partial_\nu \bm{W}_\mu
- g{\bm W}_\mu\times {\bm W}_\nu \,, \\
D_\mu \Phi &=& \left(\partial_\mu 
+ i \frac{g}{2} \bm{\tau} \cdot \bm{W}_\mu\right)\Phi \,, \\
D^\mu \bm{F}_{\mu \nu} &=& \partial^\mu \bm{F}_{\mu \nu}
- g \bm{W}^\mu \times \bm{F}_{\mu\nu} \,,
\end{eqnarray}
where $\bm{\tau}$ represents the weak isospin Pauli matrices.
We obtain the equations of motion
\begin{equation}
D_\mu \bm{F}^{\mu\nu} = \bm{J}^\nu \,, \qquad
D^\mu D_\mu \Phi = 2\lambda(\mu^2 - |\Phi|^2) \Phi\,,
\end{equation}
where the gauge current is
$\bm{J}_\nu = g {\, \rm Im \,} (D_\nu \Phi)^\dagger \bm{\tau} \Phi$
and we work in the gauge
$\bm{W}_0 = \bm{0}$. With this choice, the
covariant time derivatives become ordinary derivatives and we can
apply a Hamiltonian formalism. The $\bm{W}_j$ fields have mass
$m_W=g \mu/\sqrt{2}$, and the Higgs field has mass 
$m_H=2 \mu \sqrt{\lambda}$.

To include the effects of expansion, we again work in comoving
coordinates with a scale factor $a(t)$. We now have the action
\begin{equation}
S = \int d^3 \bm{r} \, a(t)^3 \left[
\frac{1}{2} \sum_{j=x,y,z}
\left(\bm{E}_j\cdot \bm{E}_j -
\bm{B}_j\cdot \bm{B}_j \right)
+ \dot \Phi^\dagger \dot \Phi
- \frac{1}{a(t)^2} \sum_{j=x,y,z} (\partial_j \Phi^\dagger)(
\partial_j \Phi) - \lambda(|\Phi|^2 - \mu^2)^2
\right] \,,
\end{equation}
where
\begin{equation}
\bm{E}_j = \dot{\bm{W}}_j \hbox{\quad and \quad}
\bm{B}_j =  -\frac{1}{2} \sum_{j',j''=x,y,z}
\epsilon_{jj'j''} \left(\frac{1}{a(t)} \partial_{j'} \bm{W}_{j''}
-g \bm{W}_{j'} \bm{W}_{j''}\right) \,.
\end{equation}
Here dot indicates time
derivative and Latin indices run over space dimensions.

For numerical computation we put the theory on a lattice,
following the conventions and techniques used in \cite{SU2osc3d2}. 
The field variables are the values of the $\Phi^p$ field at the
lattice sites $p$ and the spacelike Wilson lines
\begin{equation}
U_j^p(t) = e^{i g \bm{W}_j^p(t) \cdot\bm{\tau} a(t) \Delta x/2}
\label{eq:Wilson}
\end{equation}
emanating from lattice site $p$ in the spacelike direction $j$.
Since the lattice equations are second order, we will find each field
at the next time slice based on the previous two. We let $t$ be the
time for the current set of lattice points and spacelike links and
define $t_+ = t+\Delta t$  and $t_- = t-\Delta t$ to be 
the subsequent and previous times respectively. We also take $t_{+/2}
= t+\Delta t/2$ and $t_{-/2} = t-\Delta t/2$ to be the times in
between, which will be the times at which we evaluate the timelike links.

We define the Wilson line for the link emanating from
lattice site $p$ in the negative $j^{\rm th}$ direction to be the
adjoint of the corresponding Wilson line emanating in the positive
direction from the neighboring site, $U_{-j}^p(t) = 
U_j^{p-j}(t)^\dagger$,
where the notation $p \pm j$ indicates the adjacent lattice site to
$p$, displaced from $p$ in direction $\pm j$. At the edges of the
lattice we use periodic boundary conditions. We define the elements
of the field strength tensor, which are centered on the timelike and
spacelike plaquettes of the lattice,
\begin{equation}
\bm{\tau}\cdot
\bm{E}_j^p(t_{+/2}) = \frac{2}{i g a(t_{+/2}) \Delta x\Delta t}
\log U_j^p(t_+)U_j^p(t)^\dagger \hbox{\quad and \quad}
\bm{\tau}\cdot
\bm{B}_j^p(t) = \frac{i}{g (a(t) \Delta x)^2}
\sum_{j',j''=x,y, z}
\epsilon_{j j' j''} U^p_{\square(j',j'')}(t) \,,
\end{equation}
where
$U^p_{\square(j,j')}(t) = U_j^p(t)
U_{j'}^{p+j}(t) U_{-j}^{p+j+j'}(t)
U_{-j'}^{p+j'}(t)$
and  we have defined the logarithm of 
a $2\times 2$ matrix in the form of Eq.~\ref{eq:Wilson} as
\begin{equation}
\log U_j^p(t) = \frac{ig a(t) \Delta x}{2} \bm{W}_j^p(t)\cdot\bm{\tau} \,.
\label{eq:log}
\end{equation}

We note that $\log XY \neq \log X + \log Y$ when the matrices
do not commute. The logarithms and exponentials needed to
convert between the group and the algebra can be computed efficiently
using
\begin{equation}
e^{i \theta \bm{\hat n} \cdot \bm{\vec \tau}} =
\cos \theta + i \bm{\hat n}\cdot \bm{\vec \tau} \sin \theta =
\begin{pmatrix}
\cos \theta + i \bm{\hat n}_z \sin \theta &
i\bm{\hat n}_x \sin \theta + \bm{\hat n}_y \sin \theta \cr
i\bm{\hat n}_x \sin \theta - \bm{\hat n}_y \sin \theta &
\cos \theta - i \bm{\hat n}_z \sin \theta
\end{pmatrix}
\,,
\label{logformula}
\end{equation}
where $\bm{\hat n}$ is a unit vector and the link matrices have
$\bm{\hat n} \theta = \bm{W}_j^p(t) g a(t) \Delta x/2$. For efficiency
we replace $\sin\theta \rightarrow \theta$ and $\cos\theta \rightarrow
\sqrt{1- \theta^2}$ when computing both the logarithm and the
corresponding exponential. This discretization then is equivalent
(without expansion) to what is used in other numerical studies of
electroweak symmetry breaking dynamics
\cite{Shaposhnikov,RajantieSim,SmitSim,vanderMeulen}. 

We find the equation of motion for the Higgs field
\begin{equation}
\Phi^p(t_+) = 
\frac{1}{1+\frac{3H\Delta t}{2}}
\left[2 \Phi^p(t) - 
\left(1-\frac{3H\Delta t}{2}\right)\Phi^p(t_-) + \Delta t^2 \ddot
\Phi^p (t)\right] \,,
\end{equation}
where $H=\frac{\dot a(t)}{a(t)}$ is the Hubble constant and
\begin{equation}
\ddot \Phi^p (t) = \sum_{j=\pm x,\pm y, \pm z}
\frac{U_j^p(t) \Phi^{p+j}(t) - \Phi^p(t)}{a(t)^2 \Delta x^2}
 + 2\lambda\left(\mu^2 - |\Phi^p(t)|^2\right)\Phi^p(t) \,.
\end{equation}
For the gauge fields, we have
\begin{eqnarray}
U_j^p(t_+) &=&
\left(\exp\left\{\log \left(U_j^p(t) 
U_j^p(t_-)^{\frac{H\Delta t}{2} - 1}\right)
\phantom{\left(\frac{\log U^p_{\square(j,-j')}(t)}{\Delta x^2}\right)}
\right. \right. \cr && \left. \left.
-\left[\sum_{j'\neq j}  \left(
\frac{\log U^p_{\square(j,j')}(t) + 
\log U^p_{\square(j,-j')}(t)}{a(t)^2 \Delta x^2} \right)
+ \frac{ia(t) \Delta x}{2} g \bm{J}_j^p (t)\cdot \bm{\tau}
\right]\Delta t^2 \right\}
U_j^p(t)\right)^{\frac{1}{\frac{H\Delta t}{2} + 1}}\,,
\label{eq:gauge}
\end{eqnarray}
where the gauge current is
\begin{equation}
\bm{J}_j^p(t) = g {\rm \, Im \,} 
\frac{\Phi^p(t)^\dagger  \bm{\tau} U_{j}^p(t)
\Phi^{p+j}(t)}{a(t) \Delta x}
\end{equation}
and the logarithm in Eq.~\ref{eq:log} is used to compute the
exponents in Eq.~\ref{eq:gauge}.

Assuming it is obeyed by the initial conditions, time evolution
preserves the Gauss's Law constraint,
\begin{eqnarray}
\sum_{j=x, y, z}
\frac{\bm{E}_j^p(t_{+/2}) + \bm{E}_{-j}^p(t_{+/2})} {a(t_{+/2}) \Delta x}
=  \bm{J}_0^p(t_{+/2})\,,
\label{eq:Gauss}
\end{eqnarray}
where the charge density is given by
\begin{equation}
\bm{J}_0(t_{+/2}) = g {\rm \, Im \,} \left(\frac{\Phi^{p}(t_+) -
\Phi^{p}(t)} {\Delta t} \right)^\dagger \bm{\tau}
\Phi^{p}(t) \,.
\end{equation}
Energy is not conserved because in the expanding background we have
$dU = -p dV$, where $p$ is the pressure. In the lattice model we then
have
\begin{equation}
\frac{dU}{dt} = - (a(t) \Delta x)^3 H \sum_p
\left[
\frac{1}{2} \sum_{j=x,y,z}\left(\bm{E}_j^p \cdot \bm{E}_j^p +
\bm{B}_j^p \cdot \bm{B}_j^p  \right)
+ 3 |\dot \Phi|^2 - \sum_{j=x,y,z}
\left|\frac{U_j^p \Phi^{p+j} - \Phi^p}{a(t) \Delta x}\right|^2
- 3 \lambda(|\Phi^p|^2 - \mu^2)^2
\right]\,.
\end{equation}
We have verified that both the Gauss's Law and energy constraints are
well obeyed throughout our simulation.

We set initial conditions for the first two time-slices, which we
denote as $t_0$ and $t_1$, 
by occupying the modes of all the components of the Higgs and gauge
fields at temperature $T$, as in the case of a single scalar field.
Then we modify these initial conditions to make them obey Gauss's Law,
by the following steps:
\begin{itemize}
\item
First, since we have periodic boundary conditions, the total charge
should be zero. To enforce this constraint, we shift the $\Phi^p$
field on both of the first two time slices $t_0$ and $t_1$ by the same
constant,
\begin{equation}
\Phi^p \to \Phi^p -
\frac{i}{g\big|\dot{\bar{\Phi}}\big|^2}
\left(\bar{\bm{J}}_0 \cdot \bm{\tau} \right) 
\dot{\bar{\Phi}} \,,
\end{equation}
where $\bar{\bm{J}}_0$ and $\dot{\bar{\Phi}}$
are the average values of $\bm{J}_0$ and
$\dot \Phi$ over the lattice at time $t_{1/2}$, respectively.

\item
Next, we fix the longitudinal component of the gauge fields, as
described in Ref.~\cite{RajantieSim}. We take a discrete Fourier
transform of the initial charge $\bm{J}_0^p(t_{1/2})$
and gauge field $\bm{J}_j^p(t_{1/2})$ to obtain
$\widetilde{\bm{J}}_0^{\vec k}(t_{1/2})$ 
and $\widetilde{\bm{W}}_j^{\vec k}(t_{1/2})$ for the initial time step,
where $\vec k$ labels the Fourier transformed lattice. We
then modify the initial time derivative of $\bm{W}_j^p$ (by changing
its value on one of the first time slices but not the other) by sending
\begin{equation}
\dot{\widetilde{\bm{W}}}_j^{\vec k}(t_{1/2})
\to \dot{\widetilde{\bm{W}}}_j^{\vec k}(t_{1/2})
- \left(\sum_{j'=x,y,z} \vec k_{j'}
\dot{\widetilde{\bm{W}}}_{j'}^{\vec k}(t_{1/2})
+ i \widetilde{\bm{J}}_0^{\vec k}(t_{1/2})
\right) \frac{\vec{k}_j}{|\vec k|^2}
\end{equation}
and then inverting the discrete Fourier transform to obtain the
modified gauge fields, which in turn give the modified Wilson loops.
Note that we don't make any modification for $\vec k = \vec 0$, where
this transformation breaks down; that case was already handled by the
previous step.

\item
Finally, while in an Abelian theory the subtraction of the
longitudinal component would be sufficient to implement Gauss's Law,
for a nonabelian theory the nonlinear term in the field strength makes
this agreement only approximate. As a result, we adjust the phase of
$\Phi^p(t_1)$,
\begin{equation}
\Phi^p(t_1) \to 
\frac{|\Phi^p(t_1)|}{|\Phi^p(t_0)|}
{\cal U}^p(t_{1/2}) \Phi^p(t_0)
\hbox{\quad with \quad}
{\cal U}^p(t_{1/2}) = \exp\left[
-\sum_{j=x, y, z}
\frac{\log U_j^p(t_1) U_j^p(t_0)^\dagger
 + \log U_{-j}^p(t_1) U_{-j}^p(t_0)^\dagger }
{g^2 (a(t_{1/2}) \Delta x)^2   |\Phi^p(t_1)| |\Phi^p(t_0)|/2}
\right] ,
\end{equation}
leaving $\Phi^p_0(t_0)$ unchanged, in order to assure that Gauss's Law is
satisfied.

\end{itemize}

\section{SU(2) Model: Results}
\begin{figure}[htbp]
\includegraphics[width=0.49\linewidth]{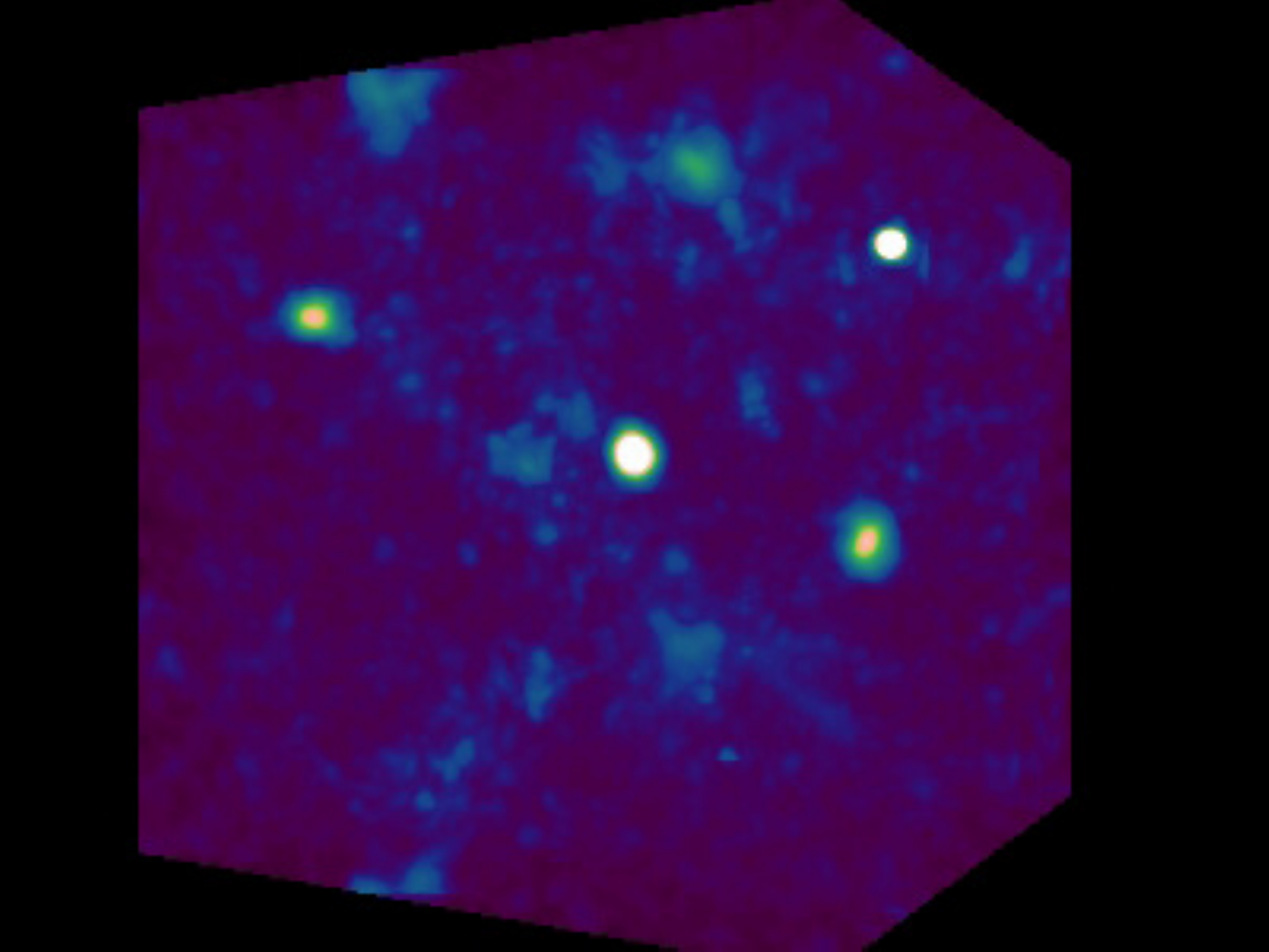}
\includegraphics[width=0.49\linewidth]{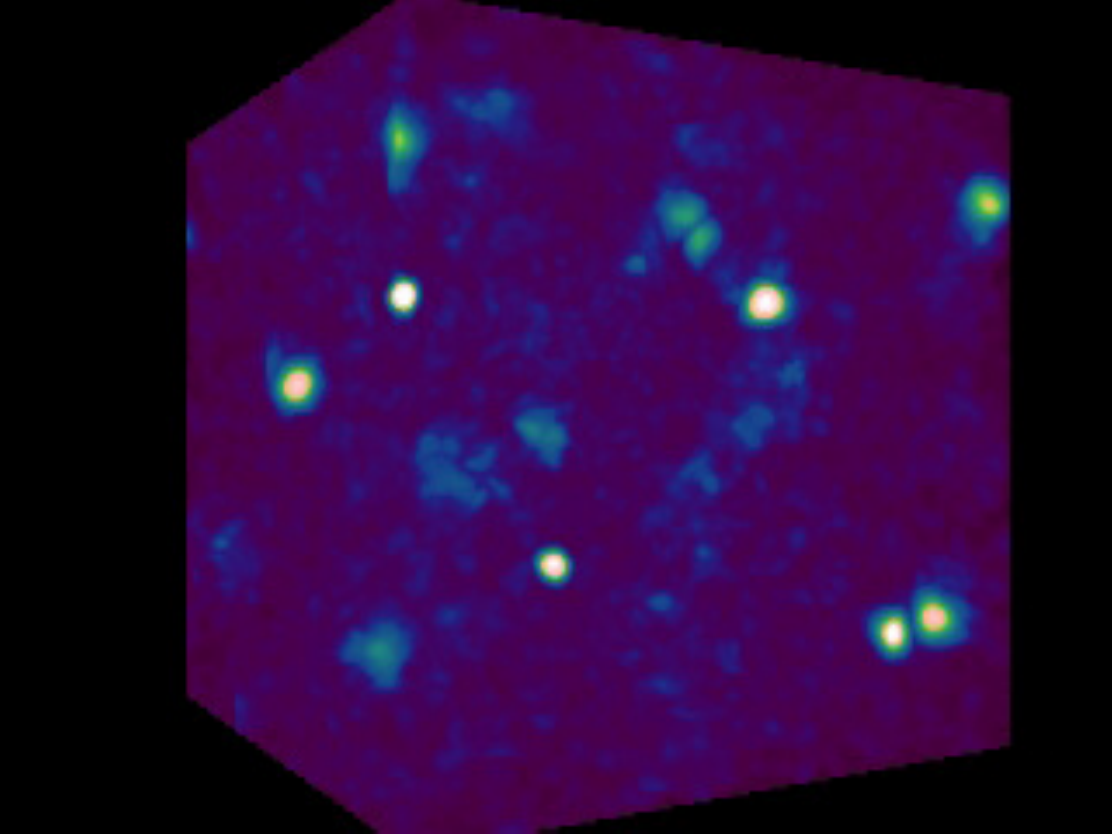}

\includegraphics[width=0.5\linewidth]{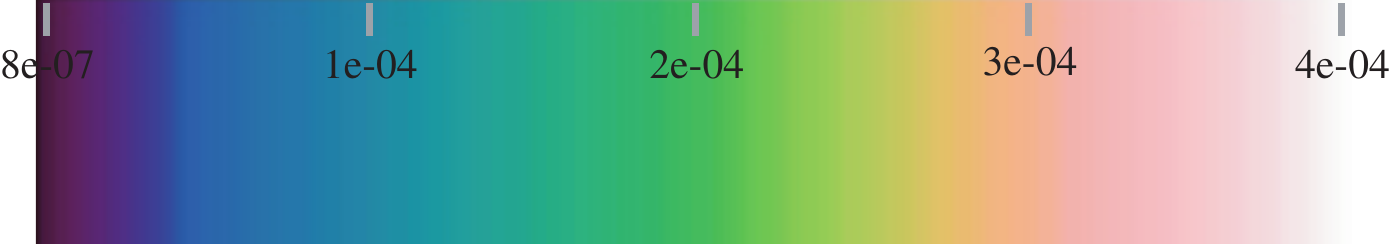}
\caption{Energy density in units of $\mu^4$ at the end of the expansion 
for $g=\sqrt{2}$ with $\lambda = 1.0$ (left panel) and $\lambda = 1.1$
(right panel). To implement the Standard Model coupling of $g_{\rm
SM}=0.624$ with the same mass ratio, this energy would be scaled up by
a factor of $(g/g_{\rm SM})^2$.}
\label{Fig:oscsu2density}
\end{figure}

\begin{figure}[htbp]
\includegraphics[width=0.49\linewidth]{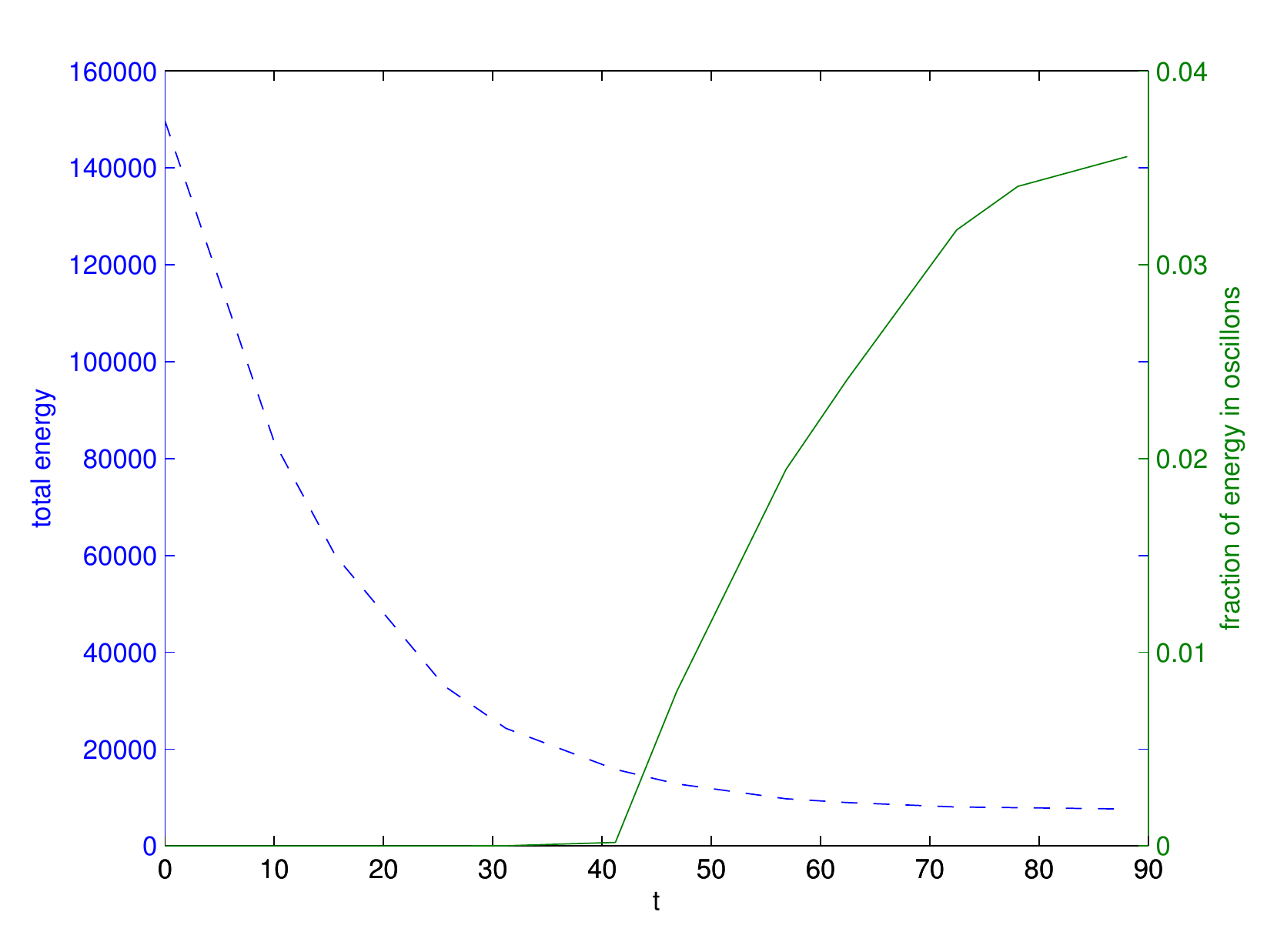}
\includegraphics[width=0.49\linewidth]{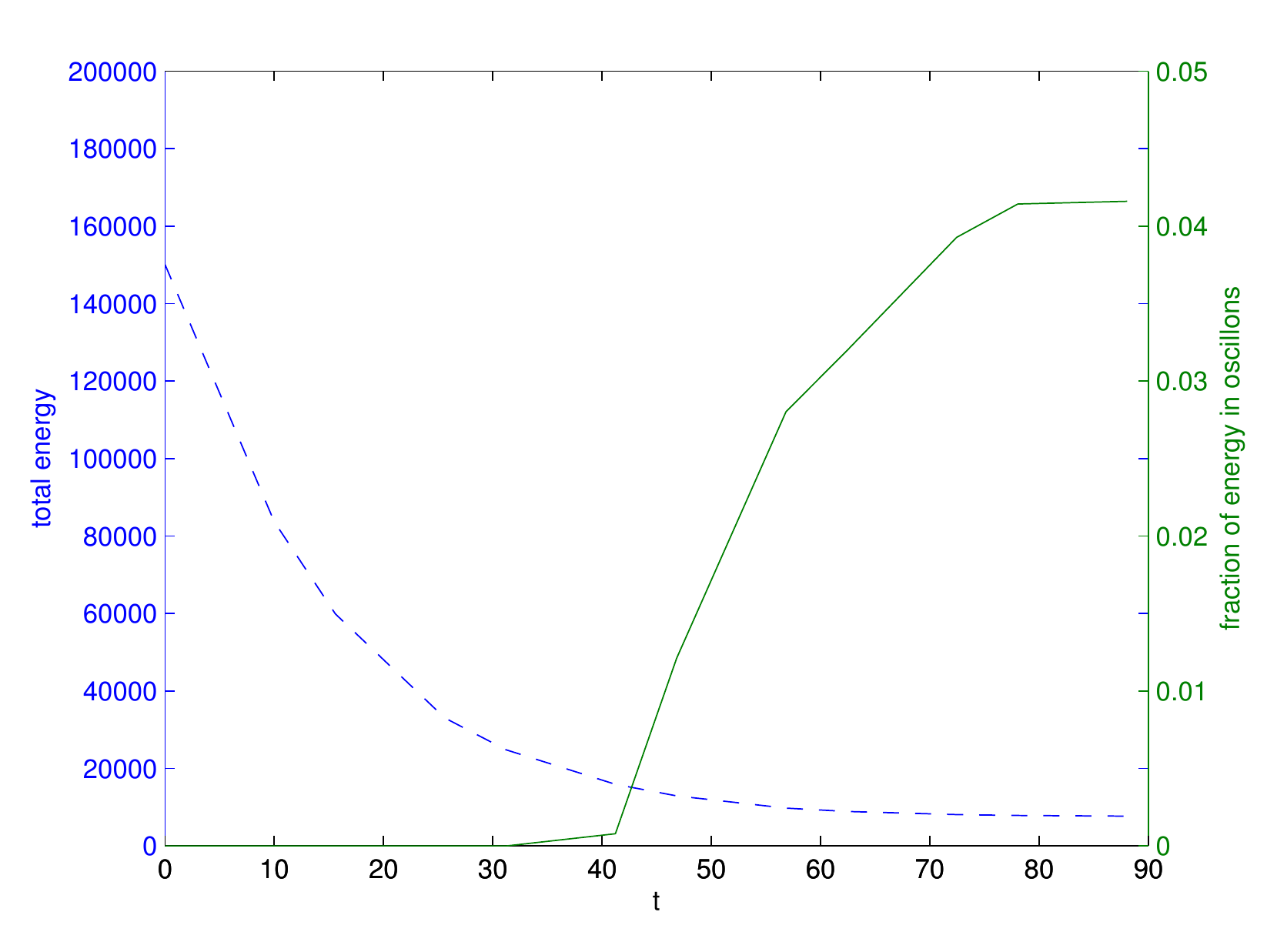}
\caption{
Fraction of energy in oscillons (solid line) and total energy (dashed
line) as functions of time. Energy is given in units of $\mu$ and time
in units of $1/\mu$, with $g=\sqrt{2}$ and $\lambda = 1.0$ (left
panel) and $\lambda = 1.1$ (right panel). To implement the Standard
Model coupling of $g_{\rm SM}=0.624$ with the same mass ratio, this
energy would be scaled up by a factor of $(g/g_{\rm SM})^2$.
}
\label{Fig:oscsu2time}
\end{figure}

We begin with a universe of size $L a(t=0)=4/\mu$, temperature $T=4 \mu$,
lattice spacing $a(t=0) \Delta x = 1/(224 \mu)$, and use a time step
$\Delta t = 1/(448 \mu)$. We allow the universe to expand at a constant
rate, with Hubble constant $H = \mu(\log 2)/12 \approx 0.06\mu$, and expand
the universe by a factor of $224$, so that the final lattice spacing is
$a(t_{\rm final}) \Delta x = 1/\mu$. We measure the fraction of
energy in oscillons by including those points whose energy density is
four times the average energy density. Under ordinary thermal
expansion this fraction would stay constant, and during the initial stages
of the expansion it is identically zero. At the end of the expansion,
approximately $4\%$ of the energy is found in oscillons by this measure.
The energy density is shown in Fig.~\ref{Fig:oscsu2density}, and the
evolution of the energy over time is shown in Fig.~\ref{Fig:oscsu2time}.
Compared to the scalar model, oscillons in the SU(2) model have
smaller amplitude and larger spatial extent, requiring a larger
simulation volume. The simulation is thus considerably more expensive
numerically, especially given the cost of evolving a total of
thirteen real degrees of freedom per lattice site instead of one.

We note that similar results are seen both for
$g=\sqrt{2}$, $\lambda=1$, in which the Higgs and gauge fields are in
the $2:1$ mass ratio found in \cite{SU2osc3d1,SU2osc3d2}, and for
$g=\sqrt{2}$, $\lambda = 1.1$, where the masses are not in this
ratio. Because of the high numerical costs associated with the
expanding background simulation, we are not able to track the 
stability of oscillons formed in this way over long time scales, but
these results suggest that the expansion may broaden the range
of parameters for which oscillons are stable. Also, by allowing
oscillons to form from a thermal background rather than a fixed
ansatz, this simulation is capable of scanning a wider
range of configuration space (and our results clearly show that
oscillon configurations represent attractors in this space).  Work is
currently underway to investigate these questions in greater detail.

\section{Possible Impact on Cosmology and Summary}
Our results indicate that oscillon-like configurations emerge
dynamically during spontaneous symmetry breaking in expanding
cosmological backgrounds. Furthermore, they contribute a significant
fraction of the total energy density. They are thus poised to play an
essential role in the dynamics of the early universe, be it during
post inflationary reheating or during symmetry-breaking phase
transitions.

In order to briefly address the impact oscillons may have on
cosmology, it is best to consider different energy scales
separately. For the sake of illustration, we focus on the GUT and
electroweak scales, which differ by roughly 13 orders of
magnitude. Also, it is important to differentiate between real scalars
and Abelian and non-Abelian Higgs models. Thus, before we start, it
may be useful to summarize what is known of oscillon lifetimes in
these models.

As we mentioned before, for real scalar fields in 3d, the oscillon
lifetime--with or without  the enhancement from the expansion reported
here--is typically $\tau_{\rm osc}\sim 10^{3-4}\mu^{-1}$
\cite{gleiser,sicilia}. For models with gauge fields, the evidence at
hand points to {\it very} large lifetimes. Studies for Abelian-Higgs
models in 2d have not seen oscillons decaying, and report lifetimes in
excess of $10^5\mu^{-1}$ \cite{U1osc2d}. Studies of Abelian-Higgs
models in 3d obtained similar results: oscillons have been observed to
persist for times $t \gtrsim 7\times 10^5\mu^{-1}$ without decaying
\cite{U1osc3d}. For non-Abelian Higgs models, the situation is
similar: the data at hand indicates that once formed, oscillons live
for extremely long times. Those in the gauged-SU(2) Higgs model have
not been observed to decay after $t \gtrsim 5\times 10^5\mu^{-1}$
\cite{SU2osc3d1,SU2osc3d2}. Thus, although a more detailed study of
Abelian and non-Abelian Higgs oscillons and their stability is clearly
warranted, results so far indicate that they may be extremely
long-lived, even perturbatively stable. Of course, the key question is
whether their lifetime can be longer than the cosmological time scale
at their formation. If that's the case, they behave as stable, localized
defects.

At the GUT scale, it is clear that the oscillon lifetime is at
least of order of the cosmological time scale, $\tilde{H}^{-1}\sim
(M_{\rm Planck}/\mu)\sim 10^{3-4}$. Thus, for all practical purposes
at GUT scales oscillons behave as stable localized defects. As has
been shown elsewhere, in the context of first-order phase transitions,
long-lived bubble-like configurations such as oscillons can either
become a critical bubble or coalesce to become one. In both cases, the
decay of the false vacuum is greatly accelerated, changing from
exponentially-suppressed to power-law \cite{vacdecay}. It
has been suggested that oscillons may accelerate the decay of the
false vacuum during inflation, potentially solving the bubble
coalescence problem of old inflation. This has been recently
illustrated within the context of a modified hybrid inflation model
\cite{oscinf}.

Oscillons may also have a key impact during post-inflationary
reheating. As coherent field configurations, they naturally delay the
approach to equilibrium, acting as bottlenecks for equipartition
\cite{gleiserhowell}. As such, they may influence (decrease) the
reheating temperature, a possibility we are currently
investigating. An interesting open question is how these
nonequilibrium results apply in the context of gauge models.

Moving on to the electroweak scale, since $\tilde{H}_{\rm ew} \sim
10^{-16}$, we are on a realm which is very distant from our numerical
range of $\tilde{H}\sim 10^{-2}$. Still, we suggest that there are at
least two ways in which oscillons may play a role at these relatively
low energy scales. Both depend on their lifetime. If non-Abelian
oscillons live for $t\sim 10^{16}\mu^{-1}$, that is, if they are
perturbatively stable, they will remain active at
cosmologically-relevant time scales. As at the GUT scale discussed
above, they may speed up vacuum decay in the context of a first-order
transition (which is ruled out in the Standard Model but not in all of
its extensions) or they may delay thermalization. If they persist for
even longer, they may even be relevant to dark matter or baryogenesis.

On the hand, if they live for shorter times $10^{4}< t \mu < 10^{16}$, their
presence may still affect the dynamics of symmetry breaking. As is
well-known, most phase transitions are initiated due to the presence
of inhomogeneities or ``seeds'' \cite{Landau}. If oscillons are
present in sufficient quantities, they will modify the effective
potential nonperturbatively, affecting the dynamics of the transition
\cite{HecklerGleiser}. Although much work remains to be done to
investigate such nonperturbative effects in more detail, these
mechanisms suggest that even relatively short-lived oscillons will
have important effects during cosmological symmetry breaking.

\noindent
{\it Acknowledgements:}
MG and NS were partially supported by a
National Science Foundation grant PHY-0653341. NG was supported in
part by the National Science Foundation (NSF) through grant
PHY-0855426 and a Baccalaureate College Development grant from Vermont
EPSCoR. Numerical simulations were carried out using TeraGrid
resources at the National Center for Supercomputing Applications
(NCSA), with support from NSF, and at the Vermont Advanced Computing
Center. Visualizations were created by David Bock (NCSA).\\

\noindent{\it Note added in proof:} A recent work by Mustafa A. Amin
offers further support to our hypothesis that oscillons will have
important effects in an expanding universe \cite{Amin}. We thank the
author for sending his manuscript to us.\\

 \end{document}